\documentclass[a4paper,aps,pra,twocolumn,groupedaddress,showkeys,showpacs,floats,floatats,floatfix]{revtex4-1}
  \usepackage{color}
  \usepackage{newlfont}
  \usepackage{graphicx}
  \usepackage{amssymb}
  \usepackage{amsmath}
  \usepackage{verbatim}
  \usepackage{bbm}
   \usepackage{mathptmx}

\usepackage[normalem]{ulem}
\usepackage{dcolumn} 
\usepackage{bm}
\usepackage{natbib}
\usepackage{latexsym}
\usepackage{mathrsfs}
\usepackage{amssymb}
\usepackage{amsmath}
\usepackage{amscd}
\usepackage[usenames,dvipsnames]{xcolor}
\usepackage{pifont}
\usepackage{verbatim}

\newcommand{\ket}[1]{\mbox{$ | #1 \rangle $}}
\newcommand{\bra}[1]{\mbox{$ \langle #1 | $}}
\newcommand{\be}{\begin{eqnarray}}
\newcommand{\ee}{\end{eqnarray}}
\begin{document}

\title{System-environment correlations for dephasing two-qubit states coupled to thermal baths}
\author{A.~C.~S.~Costa$^{1,2}$}
\author{M.~W.~Beims$^1$}
\author{W.~T.~Strunz$^2$}
\affiliation{$^1$Departamento de F\'isica, Universidade Federal do Paran\'a, 81531-980 Curitiba, Brazil\\
$^2$Institut f\"ur Theoretische Physik, Technische Universit\"at Dresden, D-01062 Dresden, Germany}

\date{\today}

\begin{abstract}
Based on the exact dynamics of a two-qubit system and environment, we investigate
{system-environment (SE)} quantum and classical correlations. The coupling is chosen 
to represent a dephasing channel
for one of the qubits and the environment is a proper thermal bath. 
First we discuss the general issue of dilation for qubit phase damping. 
Based on the usual thermal bath of harmonic oscillators, we derive criteria of separability 
and entanglement between an initial $X$ state and the environment. Applying these 
criteria to initial Werner states, we find that entanglement between the system and 
environment is built up in time for temperatures below a certain critical 
temperature $T_{\mathrm{crit}}$. 
On the other hand, the total state remains separable during those short times that are
relevant for decoherence and loss of entanglement in the two-qubit state. 
Close to $T_{\mathrm{crit}}$ the SE correlations oscillate between
separable and entangled. Even though these oscillations are also observed 
in the entanglement between the two qubits, no simple relation between the loss 
of entanglement in the two-qubit system and the build-up of entanglement between 
the system and environment is found.
\end{abstract}

\maketitle

\section{Introduction}

Quantum coherences and multipartite quantum correlations are essential 
resources for quantum information processing~\cite{Nielsen00}. However, realistic
carries of quantum information are never isolated and have to be treated as open
quantum systems~\cite{Weiss08,Breuer02}. The possibility to control and
manipulate quantum information is limited by decoherence and dissipation, usually
caused by the environment coupled to the central quantum system. Thus, a thorough
understanding of the dynamics of entanglement in open 
systems~\cite{Cubitt03,Carvalho07,Diosi03} is of fundamental importance. 
Most investigations focus on the dynamics of the reduced quantum system, tracing
over the environmental degrees of freedom. Obviously, the build-up of 
system-environment (SE) correlations can not be discussed on the reduced level.

Decoherence and loss of entanglement in an open quantum system is due to 
the build-up of SE correlations. To properly study such correlations, the
environmental degrees of freedom have to be taken into account, based
on a model for system, environment, and their 
interactions~\cite{Eisert02,Zurek09,Eisert10,Pernice11,Pernice12}.
Some work has been done in the context of total models for entangled qubit systems and 
the appearance of correlations with the surrounding environments~\cite{Lopez08,Salles08}.
There, the authors find that complete loss ({\it sudden 
death}) of entanglement in the system can manifest itself before, simultaneously, 
or even after the {\it sudden birth} of entanglement with the environment. More 
recently, entanglement in reduced bipartitions has been studied for systems 
coupled with pure environment initial states~\cite{Maziero10,Farias12,Aguilar14}.

It seems natural to assume that the loss of entanglement in an open system
is accompanied by the build-up of entanglement with the environment. One may
think of a transfer of entanglement from the local to the total state. This is
certainly true for a pure total state. However, for a finite temperature bath it
is less obvious. For realistic finite temperature baths it is a challenge
to prove entanglement or separability for the total sate.

It is an interesting question whether the SE correlations are of quantum or of classical nature. In general it is hard to answer this question for realistic
environments. Therefore, to fully understand the decoherence process, and to be
able to control the above mentioned resources, it is desirable to describe the
detailed dynamics of SE quantum correlations. One possibility is to extract, for
example, information about the SE correlations using monogamic relations for the
main system~\cite{Reina14,Costa14}. A more complete description of the full
dynamics, however, is obtained from the total state (system plus environment) by
using a coherent-basis and a partial representation of the total density
operator~\cite{Diosi97,Strunz05}. 

In addition, it is worth mentioning that SE quantum correlations 
depend on a physically appropriate dilation of qubit dephasing, where for 
different environments we obtain the same dephasing dynamics for the reduced 
system. In other words, different total systems and SE correlations lead to the 
same reduced dynamics. Therefore, in order to properly investigate SE correlations, 
it is important to choose the appropriate dilation for the total system. 

Here we are interested in a detailed study of how SE correlations
build up when decoherence and loss of entanglement occurs in the central system. 
For this we analyze the
model of two qubits, coupling one of them to an environment of dephasing
nature. This is an extension of previous investigations~\cite{Pernice11,Pernice12} 
where only one qubit was coupled to thermal baths composed by harmonic 
oscillators. In fact, since it is an analytically solvable model, we are able to
construct conditions for SE separability and entanglement, and to compare them
with the entanglement present in the two-qubit system.

This work is organized as follows. {In Sec.~\ref{dilation} we discuss the
dilation of qubit dephasing for different environments. Sections \ref{model} and 
\ref{total-state} present our model and give an exact expression for the total
state of the system. Using this expression we construct criteria for separability 
and entanglement between the system and the environment. While some results for 
distinct initial Werner states and coupling strengths are presented in 
Sec.~\ref{results}, Sec.~\ref{bipartitions} discusses entanglement within many 
different bipartitions. We reserve Sec.~\ref{conclusions}
for discussion and conclusions.}


\section{Dilating qubit dephasing}
\label{dilation}
In order to study the build-up of SE correlations, we have to specify a physical 
realization of the environment and the SE interaction. More precisely, we here are
not only interested in general correlations but in SE entanglement. This 
section serves as a simple introduction to clarify the relevance of the choice 
of dilation when studying SE correlations, in particular for mixed
environmental initial states. In any case, we need to determine
the total SE state. 
 
 Locally, within the framework of 
completely positive (CP) and trace preserving maps, single-qubit dephasing in 
the computational basis is given by the quantum channel, \cite{Nielsen00}
\begin{equation}\label{qubitdephasing}
 \rho\rightarrow\rho'= {\cal E}[\rho] = \frac{1+\sqrt{p}}{2}\,\rho + \frac{1-\sqrt{p}}{2}\,
 (\sigma_3\rho\sigma_3),
\end{equation}
with the third Pauli matrix $\sigma_3$, or, in matrix notation,
\begin{equation}
 \rho = \left(\begin{array}{cc}\rho_{00} & \rho_{01} \\  \rho_{10} & \rho_{11}\end{array}\right) 
 \rightarrow 
 \rho' = \left(\begin{array}{cc}\rho_{00} & \sqrt{p}\, \rho_{01} \\  
\sqrt{p}\,\rho_{10} & \rho_{11}\end{array}\right).
\end{equation}
Here, the real $p$ with $0\le p \le 1$ takes the role of the dephasing parameter: $p=1$ corresponds
to no dephasing, while
$p=0$ describes the full loss of coherence. Often, $p$ will be some (decaying)
function of time, depending on coupling strength and, for a thermal bath, on the
temperature of the environment (see later). 

Before discussing various dilations of the dephasing channel (\ref{qubitdephasing}), 
it is worth noting that all unital single-qubit channels (including dephasing) are
of so-called random-unitary (RU) type, i.e., they can be obtained from an ensemble
of unitary evolutions without invoking a quantum environment at all~\cite{Landau93}. 
It is only for two-qubit systems (and larger) that dephasing may be of true
quantum nature \cite{Helm09,Helm10}.

Thus, the local point of view (\ref{qubitdephasing}) does not allow for any
conclusions about SE correlations. We need to specify the underlying total
dynamics ($U_{\mathrm{tot}}$) and the environmental initial state ($\rho_E$).  Then,
the dilation,
\begin{equation}
 \rho' = {\cal E}[\rho] = {\textrm{Tr}}_\mathrm{E}\left(U_{\mathrm{tot}}(\rho\otimes\rho_\mathrm{E})U^\dagger_{\mathrm{tot}})\right),
\end{equation}
allows us to study SE correlations. Clearly, depending on the choice of dilation,
different SE correlation scenarios are possible, as we will point out next.

For pure qubit dephasing studied here, without loss of generality the total
unitary evolution can be written in the form \cite{Roszak15},
\begin{equation}\label{totalU}
 U_\mathrm{tot}=|0\rangle\langle 0|\otimes 1_\mathrm{E} + |1\rangle\langle 1|
\otimes U_\mathrm{E},
\end{equation}
with the two ``open system'' qubit states $|0\rangle$ and $|1\rangle$ and a
unitary evolution operator $U_\mathrm{E}$ of the environment, conditioned on the
qubit state $|1\rangle$. Thus, any dilation of qubit dephasing is fully determined
by the initial state $\rho_\mathrm{E}$ and a unitary evolution operator 
$U_\mathrm{E}$ of the environment.
As we will see, for the build-up of SE entanglement, the purity of the 
environmental initial state is of great relevance. 

\subsection{Pure environmental initial state: entangling dilation}
Often, a {\it pure} environmental state $\rho_\mathrm{E}=|0_\mathrm{E}\rangle
\langle 0_\mathrm{E}|$ is assumed~\cite{Maziero10, Farias12, Aguilar14, Roszak15}. 
Then, by construction, the total state dynamics is entirely determined from the 
two equations,
\begin{eqnarray}
U_\mathrm{tot} |0\rangle|0_\mathrm{E}\rangle & = & |0\rangle|0_\mathrm{E}\rangle,
\\ \nonumber
U_\mathrm{tot} |1\rangle|0_\mathrm{E}\rangle & = & 
|1\rangle \left(\sqrt{p}\,|0_\mathrm{E}\rangle+\sqrt{1-p}\,|1_\mathrm{E}\rangle
\right),
\end{eqnarray}
where the relation,
\begin{equation}\label{envone}
  U_\mathrm{E}|0_\mathrm{E}\rangle = \sqrt{p}\,|0_\mathrm{E}\rangle+\sqrt{1-p}\,|1_\mathrm{E}\rangle,
\end{equation}
defines $p$ and the environmental state $|1_\mathrm{E}\rangle$ on the right-hand 
side of the equation (we neglect a possible, yet irrelevant phase here). 
Thus, for a {\it pure} environmental initial state $|0_\mathrm{E}\rangle$, only 
one orthogonal environmental state $|1_\mathrm{E}\rangle$ [as defined in 
(\ref{envone})] is relevant and the true environment can be effectively described
by a single qubit, as in \cite{Maziero10, Farias12, Aguilar14}. This single-qubit
environment dilation is thus defined by the two choices,
\begin{eqnarray}\label{pure_dilation}
 \rho_\mathrm{E} & = & |0_\mathrm{E}\rangle\langle 0_\mathrm{E}|, \\ \nonumber
 U_\mathrm{E} & = & \sqrt{p}\,\,\sigma_3 + \sqrt{1-p}\,\,\sigma_1.
\end{eqnarray}

It is easy to show that the partial transpose of the corresponding effective two-qubit SE state has a determinant of 
\begin{equation}
 \det\{\rho_\mathrm{tot}^\mathrm{PT}\}= - (1-p)^2\rho_{00}\rho_{11}|\rho_{01}|^2.
\end{equation}
Using the Peres criterion \cite{Peres96} we conclude that starting from a pure
environmental initial state $|0_\mathrm{E}\rangle$, the effective SE two-qubit
state will be entangled for both, $p<1$ and the initial $\rho_{01}\neq 0$, i.e., 
whenever some initial coherence is present and dephasing actually happens.

\subsection{Mixed environmental initial state: separable dilation}
Such SE entanglement need not develop for a {\it mixed} environmental initial
state~\cite{Pernice11, Pernice12}. To give a simple example, consider again a 
single-qubit dilation of the dephasing channel (\ref{qubitdephasing}), now with
\begin{eqnarray}\label{mixed_dilation}
 \rho_\mathrm{E} & = & \frac{1+\sqrt{p}}{2}|0_\mathrm{E}\rangle\langle 0_\mathrm{E}|
 + \frac{1-\sqrt{p}}{2}|1_\mathrm{E}\rangle\langle 1_\mathrm{E}|, \\ \nonumber
 U_\mathrm{E} & = & \sigma_3.
\end{eqnarray}
In contrast to the pure dilation (\ref{pure_dilation}), where the parameter $p$ 
may be interpreted as representing time, here $p$ should be rather interpreted as
a measure for initial environmental temperature (large $p\rightarrow 1$
corresponding to low $T\rightarrow 0$ and vice versa) and dynamics evolves for a
fixed time. As with (\ref{pure_dilation}), it is easily checked that 
(\ref{mixed_dilation}) is a valid dilation of pure dephasing, leading to the CP
map (\ref{qubitdephasing}) for the reduced state of the qubit.
Now, using (\ref{totalU}), and in contrast to the previous dilation based on 
(\ref{pure_dilation}), the total state is separable for all $p$,
\begin{equation}
 \rho_\mathrm{tot} = \frac{1+\sqrt{p}}{2}\rho\otimes |0_\mathrm{E}\rangle\langle 0_\mathrm{E}|
 + \frac{1-\sqrt{p}}{2}(\sigma_3\rho\sigma_3)\otimes |1_\mathrm{E}\rangle\langle 1_\mathrm{E}|,
\end{equation}
i.e., no SE entanglement builds up.

We conclude from these considerations that in order to study SE entanglement, it
is crucial to choose a physically appropriate dilation. Matters are considerably
more involved for a mixed environmental initial state: Even for dephasing we can 
no longer expect to describe the environment by an effective qubit.
Indeed, the action of $U_\mathrm{tot}$ [or rather $U_\mathrm{E}$ in 
(\ref{totalU})] is no longer restricted to a single initial environmental state 
$|0_\mathrm{E}\rangle$. Thus, in general, the dynamically relevant environmental Hilbert space
can no longer be spanned by just two states as in (\ref{envone}).

In the following, we study a {\it two-qubit} system, coupled to a proper
environment consisting of an infinite number of degrees of freedom and initially
in a thermal (mixed) state. As the total state is neither a two-qubit nor a
Gaussian state, the detection of entanglement is a nontrivial issue
and will be based on the negativity of the partial transpose \cite{Peres96}.


\section{Model for system and environment}
\label{model}

\begin{figure}
\includegraphics[scale=0.35]{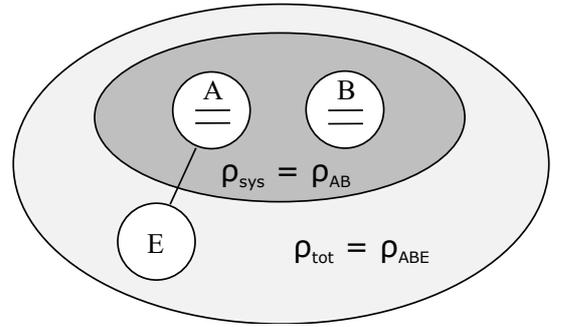}
\caption{Scheme of the proposed model. System of two qubits ($AB$) with qubit $A$ coupled to the environment ($E$).}
\label{Fig1}
\end{figure}

We study an initially entangled two-qubit (qubits $A$ and $B$) state. One of the
qubits ($A$) is coupled to its (local) environment---representing a single-qubit
dephasing channel for qubit $A$. In Fig.~\ref{Fig1}, we can see a scheme of the underlying model. The environment is chosen to be a bath of
harmonic oscillators, initially in a thermal state, a standard model for open
quantum system dynamics~\cite{Feynman63,Caldeira81}. As we aim at pure dephasing,
the Hamiltonian of the system and the system part of the coupling of qubit $A$ to 
its environment have to be diagonal in the chosen (computational) basis. Thus, for
the total Hamiltonian we write
$H_{\mathrm{tot}} = H_{\mathrm{sys}} + H_{\mathrm{int}} + H_{\mathrm{env}}$, 
with a system Hamiltonian,
\be
H_{\mathrm{sys}} = \frac{\hbar\,\Omega_A}{2}(\sigma_3^A\otimes\mathbbm{1}^B)
+\frac{\hbar\,\Omega_B}{2}(\mathbbm{1}^A\otimes\sigma_3^B),
\ee 
a (bosonic) bath of harmonic oscillators (labelled by $\lambda$ and creation
and annihilation operators $a_\lambda, a^\dagger_\lambda$
with commutation relations $[a_\lambda,a^\dagger_\mu]=\delta_{\lambda\mu}$), 
$H_{\mathrm{env}} = \sum_{\lambda} \hbar\omega_\lambda a_\lambda^\dagger a_\lambda$, 
and diagonal (with respect to the system) interaction between qubit $A$ and 
the environment of the form
\be
H_{\mathrm{int}} = (\sigma_3^A\otimes\mathbbm{1}^B)
\otimes\sum_{\lambda} \hbar(g_\lambda^\ast a_\lambda^\dagger 
+ g_\lambda a_\lambda).
\ee
$\Omega_{A,B}$ are the characteristic frequencies of the qubits and the
coefficients $g_\lambda$ are the coupling amplitudes between the qubit $A$ 
and each environmental mode of frequency $\omega_\lambda$. The full
SE Hamiltonian underlying our investigations represents a standard model
of quantum dephasing of qubits and can be found in numerous earlier works---see, for instance,~\cite{Palma96,Reina02,Brito15}.

We assume that the environment is initially in a thermal state at temperature $T$,
expressed by the canonical density operator,
\be
\rho_{\mathrm{therm}} = \frac{1}{Z}\exp (-H_{\mathrm{env}}/k_B T),
\ee
with partition function $Z={\textrm{Tr}[\exp (- H_{\mathrm{env}}/k_B T)]}$. The 
mean thermal occupation number is the usual $\bar n_\lambda = 
(\exp [\hbar\omega_\lambda/k_B T] - 1)^{-1}$. 
The total initial state is the product,
\begin{equation}\label{irhot}
\rho_{\mathrm{tot}} (0) = \rho_{\mathrm{sys}} (0) \otimes \rho_{\mathrm{therm}},   
\end{equation}
with a (possibly entangled) initial two-qubit state
$\rho_{\mathrm{sys}} (0) $. 

In fact, in what follows, due to their analytical tractability, we choose the so-called $X$ states~\cite{Yu07}. Thus, the two-qubit initial state in the 
computational basis is given by the matrix,
\be\label{xstate}
\rho_{\mathrm{sys}} (0) = \left(
\begin{array}{cccc}
\rho_{11} & 0 & 0 & \rho_{14} \\
0 & \rho_{22} & \rho_{23} & 0 \\
0 & \rho_{23}^\ast & \rho_{33} & 0 \\
\rho_{14}^\ast & 0 & 0 & \rho_{44}
\end{array}
\right),
\ee
where $\sum_{i=1}^4 \rho_{ii} = 1$. The family of $X$ states includes pure Bell 
states and the well-known Werner states~\cite{Werner89}. Crucially, the family of
$X$ states is closed under dephasing dynamics.

The chosen model allows us to derive an exact master equation \cite{Pernice11} for 
the reduced density operator $\rho_{\mathrm{sys}} (t) = \textrm{Tr}_{\mathrm{env}} 
[\rho_{\mathrm{tot}} (t)]$ 
which reads 
\be\label{master-eq}
\dot \rho_{\mathrm{red}} &=& - \frac{i\,\Omega_A}{2}[\sigma_3^A\otimes\mathbbm{1}^B,\rho_{\mathrm{red}}] -
\frac{i\,\Omega_B}{2}[\mathbbm{1}^A\otimes\sigma_3^B,\rho_{\mathrm{red}}] \nonumber \\ 
&& - \frac{\gamma_{dph}
(t)}{2}\Big(\rho_{\mathrm{red}} -
(\sigma_3^A\otimes\mathbbm{1}^B)\rho_{\mathrm{red}}(\sigma_3^A\otimes\mathbbm{1}^B)\Big).\quad
\ee
Equation (\ref{master-eq}) takes the form of a master equation of Lindblad type with,
however, a time-dependent dephasing rate $\gamma_{dph}(t) \equiv \gamma$.
Indeed, in terms of the environmental spectral density 
$J(\omega) = \sum_{\lambda} |g_\lambda|^2\delta (\omega -\omega_\lambda)$, the 
dephasing rate is given by
\be\label{gamma}
\gamma_{dph} (t) = 4\int_0^t ds\int_0^\infty d\omega
J(\omega)\coth\left[\frac{\hbar\omega}{2 k_B T}\right] \cos[\omega s].\qquad
\ee
Note that this rate may turn negative at times for nontrivial spectral densities 
and therefore the CP map $\rho(0)\to\rho(t)$ may lose its
divisibility, which is used as an indication for non-Markovian quantum
dynamics~\cite{Pernice12,WolfCirac08,Rivas10}.

The solution of Eq.~(\ref{master-eq}) with initial state (\ref{xstate}) is the
reduced state,
\be\label{rho-red}
\rho_{\mathrm{red}} (t) = \left(\begin{array}{cccc}
\rho_{11} & 0 & 0 & \rho_{14}\mathcal{D}_{+}(t) \\
0 & \rho_{22} & \rho_{23}\mathcal{D}_{-}(t) & 0 \\
0 & \rho_{23}^\ast\mathcal{D}_{-}^\ast(t) & \rho_{33} & 0 \\
\rho_{14}^\ast\mathcal{D}_{+}^\ast(t) & 0 & 0 & \rho_{44}
\end{array}
\right), \nonumber \\
\ee
where $\mathcal{D}_{\pm}(t) = \exp \left[i(\Omega_A \pm \Omega_B) - \int_0^t 
\gamma_{dph}(s)ds\right]$. 

Entanglement within the two-qubit system can be calculated via 
{\sl concurrence}~\cite{Wootters98}, 
a well-known measure for mixed two-qubit states. For our model this measure of
entanglement is given by
\be\label{x_con}
\mathcal{C}(\rho_{\mathrm{red}} (t)) = 
2\max \Big\{0,&&|\rho_{23}||\mathcal{D}(t)| - \sqrt{\rho_{11}\rho_{44}},\nonumber \\
&& |\rho_{14}||\mathcal{D}(t)|-\sqrt{\rho_{22}\rho_{33}}\Big\},
\ee
where $|\mathcal{D}(t)|=|\mathcal{D}_{\pm}(t)|$. Concurrence varies between 0
(separable states) and $1$ (maximally entangled states). Clearly, due to the
dephasing dynamics, $\mathcal{C}$ typically decreases with time. Moreover,
as is apparent from (\ref{x_con}), entanglement may disappear entirely, even for
a finite time (finite $|\mathcal{D}(t)|$)---sometimes referred to as 
{\it sudden death}~\cite{Yu07}.

On the reduced level, the loss of entanglement in a two-qubit state due to a local
dephasing channel has been studied in many publications~\cite{Konrad08,Yu07, Ficek06}. 
That loss is accompanied by the build-up of correlations between the open
quantum system and its environment. In order to study whether initial entanglement 
within the open two-qubit state just disappears or whether it is transferred to 
entanglement between system and environment (or to entanglement within the 
elements of the environment), we need to determine the total state.

\section{Total state dynamics}
\label{total-state}

The study of correlations between the system and the infinite oscillator 
environment requires an expression for the total state. For this purpose we use a
coherent state basis for the environmental degrees of freedom and choose a
partial $P$ representation~\cite{Strunz05}. We here
follow closely a similar analysis for the dephasing of a single qubit presented in~\cite{Pernice11, Pernice12}. 
The total state is written as
\be\label{rhot}
\rho_{\mathrm{tot}}(t) = \int \frac{d^2 z}{\pi}\frac{1}{\bar n}e^{-|z|^2/\bar n} \hat P(t;z,z^\ast)\otimes \ket{z}\bra{z},
\ee
where $z = (z_1,z_2,...)$ is a vector of complex numbers representing 
environmental coherent state labels and we use the notation 
$d^2 z/\pi = d^2 z_1/\pi d^2 z_2/\pi ...$, and 
$\exp [-|z|^2/\bar n] = \prod_\lambda \exp[-|z_\lambda|^2/\bar n_\lambda]/\bar n_\lambda$.
Note that for $t=0$, we have the initial $\hat P(t=0) = \rho_\mathrm{sys}(0) = \rho_\mathrm{red}(0)$
such that (\ref{rhot}) represents the factored state (\ref{irhot}). It is only for 
$t>0$ that $P(t)$ becomes $z$-dependent and thus, expression (\ref{rhot})
represents a correlated SE-state.

Expression~(\ref{rhot}) is a solution of the total von Neumann equation. We find
the time evolution of the partial $P$ function to be given by
\be
\hat P(t;z,z^\ast) = \left(
\begin{array}{cccc}
\mathcal{A}_+\rho_{11} & 0 & 0 & \mathcal{B}_+\rho_{14} \\
0 & \mathcal{A}_+\rho_{22} & \mathcal{B}_-\rho_{23} & 0 \\
0 & \mathcal{B}_-^{\ast}\rho_{23}^\ast & \mathcal{A}_-\rho_{33} & 0 \\
\mathcal{B}_+^{\ast}\rho_{14}^\ast & 0 & 0 & \mathcal{A}_-\rho_{44}
\end{array}
\right),\nonumber
\ee
where $\mathcal{A}_{\pm}(t) = \exp [-\alpha (t)\pm\{(a(t)|z)+(z|a(t))\}]$ and 
$\mathcal{B}_{\pm}(t) = e^{-i(\Omega_A \pm \Omega_B)t}\exp[\beta (t)-\{(b(t)|z)-(z|b(t))\}]$. 
Here $a(t) = (a_1(t),a_2(t),...)$ 
and $b(t)$ are complex time-dependent vectors,
\be
&a_\lambda (t) = \frac{1}{\bar n_\lambda}\int_0^t (g_\lambda e^{i\omega_\lambda s})ds,& \nonumber \\
& &\nonumber\\
&b_\lambda (t) = \frac{2\bar n_\lambda +1}{\bar n_\lambda}\int_0^t (g_\lambda e^{i\omega_\lambda s})ds,&
\ee
with the scalar product $(a(t)|z) \equiv \sum_\lambda a^\ast_\lambda (t)z_\lambda$ and
\be
\alpha (t) &=& 2\Re \int_0^t ds \int_0^s d\tau \left[\sum_\lambda\frac{1}{\bar n_\lambda}|g_\lambda|^2 e^{-i\omega_\lambda (s-\tau)}\right],\nonumber \\
\beta (t) &=& 2\Re \int_0^t ds \int_0^s d\tau \left[\sum_\lambda\frac{2\bar n_\lambda +1}{\bar n_\lambda}|g_\lambda|^2 e^{-i\omega_\lambda (s-\tau)}\right]. \nonumber
\ee

\subsection{System-environment separability}

Expression (\ref{rhot}) allows us to study correlations between system and 
environment. In particular, it is clear that it is a {\it separable}
representation of the total state whenever $\hat P$ is a positive two-qubit matrix
\cite{Pernice11}. 

{\sl Separability criterion:} As long as the partial $P$-function is positive 
semidefinite, the total state $\rho_{\mathrm{tot}} (t)$ in representation
(\ref{rhot}) is trivially separable. Being of $X$ type, the eigenvalues can be 
determined analytically. Initially, all eigenvalues are positive and they remain 
positive as long as
\be\label{sep}
\mathcal{S} (t) \leq \frac{1}{2}\ln \left(\frac{\rho_{22}\rho_{33}}{|\rho_{23}|^2}\right) \,\, \textrm{and} \,\, 
\mathcal{S} (t) \leq \frac{1}{2}\ln \left(\frac{\rho_{11}\rho_{44}}{|\rho_{14}|^2}\right), \,\,\,\,\,\,\,\,
\ee
where we have defined
\be\label{sep_function}
\mathcal{S}(T,t) &:=& \alpha(t) + \beta(t) 
\nonumber \\
&=& 4\int_0^t ds\int_0^s d\tau \int_0^\infty d\omega \nonumber \\
&& \times J(\omega)e^{\hbar\omega /k_BT}\cos [\omega (s - \tau)].
\ee
Depending on the choice of the spectral density $J(\omega)$, Eq.~(\ref{sep_function}) 
can be written in terms of known special functions. Thus,
as long as (\ref{sep}) is satisfied (as a function of time and temperature), the
total state is separable and no SE entanglement builds up, even though the initial
two-qubit $X$ state may well lose its initial entanglement (see results later).

\subsection{System-environment entanglement}

The detection of entanglement between system and environment is very demanding 
since the underlying state is infinite-dimensional and not of Gaussian type. We
can use the Peres criterion~\cite{Peres96} to see that entanglement is there,
surely, if the partial transpose 
$\rho^{\mathrm{PT}}_{\mathrm{tot}}(t)$ of the total state has 
a negative expectation value 
$\varepsilon^{\mathrm{PT}} = \bra{\Psi}\rho^{\mathrm{PT}}_{\mathrm{tot}}\ket{\Psi}$ 
for some suitably chosen total system state $\ket{\Psi}$.

We expand $\ket{\Psi}=\int \frac{d^2z}{\pi} e^{-|z|^2}|\psi(z^*)\rangle ||z\rangle$ 
in a Bargmann coherent state basis. Thus, in order to detect entanglement, we need
to find a system state $\ket{\psi (z^\ast)} \sim \langle z\ket{\Psi
(t)}$ analytical in $z^*$ in the Hilbert space of the two qubits such that
\be
\varepsilon^{\mathrm{PT}} \sim \int \frac{d^2 z}{\pi}e^{-\frac{\bar n+1}{\bar n}|z|^2} 
\bra{\psi (z)}\hat P^T(z,z^\ast)\ket{\psi (z^\ast)} < 0. \quad
\ee
After some experimenting we choose
\be\label{teststate}
\ket{\psi (t,z^\ast)} = \left( 
\begin{array}{c}
r\, e^{-(z|a+b)/2 + i (\Omega_A + \Omega_B)t/2} \\
-s\, e^{-(z|a+b)/2 + i (\Omega_A - \Omega_B)t/2} \\
-t\, e^{(z|a+b)/2 - i (\Omega_A - \Omega_B)t/2} \\
u\, e^{(z|a+b)/2 - i (\Omega_A + \Omega_B)t/2}
\end{array}
\right).
\ee
Here we found that for optimal entanglement detection, the vector $(r,s,t,u)$
needs to be determined as the pure state which has the smallest overlap with the
transpose of the initial state of the two-qubit system.

Performing the integral over the coherent state labels $z_\lambda$, we find
\be\label{ent_PT}
\varepsilon^{\mathrm{PT}} (t) &=& \frac{1}{\bar n + 1}\Big[e^{-\alpha (t) + \frac{\bar{\mathcal{S}}(t)}{2}}(\rho_{11}|r|^2 + \rho_{22}|s|^2 + \rho_{33}|t|^2 \nonumber \\
&& + \rho_{44}|u|^2) + e^{\beta (t) - \frac{\bar{\mathcal{S}}(t)}{2}}(\rho_{23}s t^\ast + \rho_{23}^\ast s^\ast t + \rho_{14} r u^\ast \nonumber \\
&& + \rho_{14}^\ast r^\ast u)\Big],
\ee
where $\bar{\mathcal{S}}(t)$ is defined similar to  $\mathcal{S}(t)$ in~(\ref{sep}), 
but with $\exp(\hbar\omega/kT)$ replaced by its inverse, $\exp(-\hbar\omega/kT)$.

From (\ref{ent_PT}) we conclude that SE entanglement is surely present whenever
the following condition is satisfied:
\be\label{ent}
\mathcal{E}(T,t) &=& \mathcal{S}(T,t) - \bar{\mathcal{S}}(T,t) \\ 
&>& \ln \left[- \frac{\rho_{11}|r|^2 + \rho_{22}|s|^2 + \rho_{33}|t|^2 + 
\rho_{44}|u|^2}{\rho_{23}s t^\ast + \rho_{23}^\ast s^\ast t + \rho_{14}r u^\ast + 
\rho_{14}^\ast r^\ast u}\right].\qquad
\nonumber
\ee

The relevant quantity on the l.h.s.~of (\ref{ent}) can be written in terms of the 
spectral density as
\be
\mathcal{E}(T,t) &=& 8\int_0^t ds\int_0^s d\tau \int_0^\infty d\omega \nonumber \\
&& \times J(\omega)\sinh (\hbar \omega /k_BT)\cos [\omega(s-\tau)].
\ee
Thus, with (\ref{ent}) we found  a criterion that allows us to detect SE entanglement
as a function of time and temperature of the bath.

In the next section we will show the regions of SE separability and entanglement
in the temperature-time diagram $(T,t)$, defined by conditions (\ref{sep}) for 
separability, and (\ref{ent}) for entanglement. Let us already mention at this 
point that these two conditions do not fill the whole ($T,t$) plane. There will be $(T,t)$
combinations where we cannot make any statement about whether the total state is
entangled or separable. This is due to the fact that our criteria are sufficient,
but not necessary conditions. For entanglement on one hand, even if our
test state (\ref{teststate}) were optimal, there may well be entangled states with
positive partial transpose (``bound entangled states'' \cite{Horodecki98}).
For separability, on the other hand, there may well be separable states with a
negative partial $\hat P$ representation. 

\section{Numerical results for Werner states}
\label{results}

In this Section we present and discuss concrete results concerning entanglement and 
separability of the total SE state. Choosing an Ohmic spectral density with a 
cutoff frequency $\omega_c$, with $J(\omega) = \kappa \omega \Theta (\omega - 
\omega_c)$, where $\kappa$ is the coupling strength between system and environment,
we can construct diagrams of separability and entanglement for different 
initial states and coupling strengths (varying~$\kappa$).

Furthermore, we want to compare the time for the decay of initial entanglement and 
the time for the build-up of SE entanglement with the typical decoherence time of
the open two-qubit system. The latter time, $\tau_\mathrm{dec}$, we define through
the relation $\int_0^{\tau_\mathrm{dec}}\gamma_{dph}ds = 1$, which defines the time
scale of the decay of the off-diagonal elements of the reduced operator in Eq.~(\ref{rho-red}).

As initial two-qubit X-states we choose Werner states. These are a family of states 
that depend on a single purity parameter $c$, and are given by $\rho^W = 
\frac{(1-c)}{4}\mathbbm{1}^{AB} + c\ket{\phi^-}\bra{\phi^-}$, where 
$\mathbbm{1}^{AB}$ is the identity matrix in the Hilbert space of the two qubits
and the Bell state $\ket{\phi^-} = (\ket{01}-\ket{10})/\sqrt{2}$, with 
$c\in [0,1]$. This state is entangled if $c > 1/3$. 

For this specific state the condition for separability, from Eq.~(\ref{sep}), 
is
\be\label{sep-w}
\mathcal{S}(T,t) \leq \ln \left[\frac{1+c}{2c}\right],
\ee
and for entanglement, from Eq.~(\ref{ent}), is
\be\label{ent-w}
\mathcal{E} (T,t) > \ln \left[\frac{1+c}{2c}\right].
\ee
The pure state that has the smallest overlap with the transpose of the two-qubit system initial state is 
\be
(r,s,t,u) = \left(0,\frac{1}{\sqrt{2}},\frac{1}{\sqrt{2}},0\right).
\ee

In the following we discuss weak SE interaction first ($\kappa=10^{-3}$),
followed by strong coupling ($\kappa = 1$). In both cases we choose three 
different initial two-qubit states: $c=0.2$ (no initial two-qubit entanglement, 
low purity), $c=0.5$ (some initial entanglement, medium purity), and $c=0.9$ 
(large initial entanglement, fairly pure).  In all cases we see an initial phase where the total 
state remains separable. For very low temperatures this phase is hardly 
visible and SE entanglement builds up quickly. We also display the loss of initial 
two-qubit entanglement and initial two-qubit coherence that is observed in all cases.
Numerical results are presented in temperature-time ($T,t$) diagrams 
following the notation: red (black gray) color when the entanglement condition 
[Eq.~(\ref{ent-w})] is satisfied and blue (light gray) color when the separability 
condition [Eq.~(\ref{sep-w})] is satisfied.

\subsection{\sl Weak coupling ($\kappa = 10^{-3}$)}

\begin{figure}[htb] 
\includegraphics[scale=0.34]{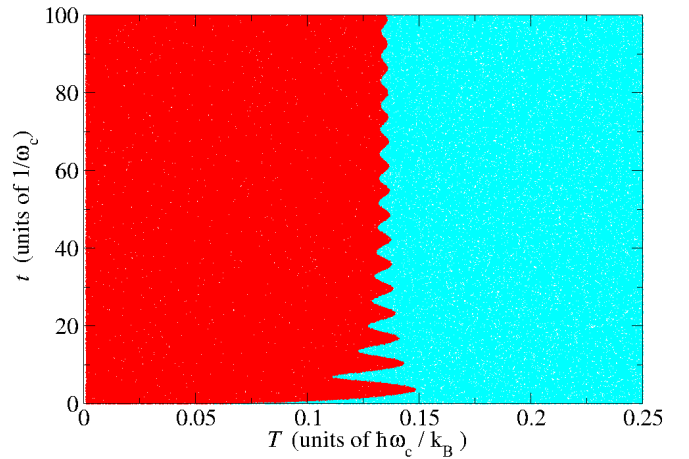} 
\caption{(Color online) Temperature-time diagram. Red (dark gray) region: 
entangled SE state, blue (light gray) region: separable SE state. Parameters 
are $\kappa = 10^{-3}$ (weak coupling) and $c=0.2$ (no initial entanglement between 
the two qubits). Remarkably, for $k_BT_{\mathrm{crit}}\approx 0.13\hbar \omega_c$, 
the total state oscillates as a function of time between separable and entangled 
regions.}
\label{wc-2} 
\end{figure}

We start with
an initial Werner state with parameter $c=0.2$. There is never any entanglement
between the two qubits, but we observe from Fig.~\ref{wc-2} that for low
temperatures, due to SE interaction, entanglement between the two-qubit 
system and environment builds up [red (dark gray) region]. There is a clearly
visible boundary at a {\it critical temperature}  
$k_BT_{\mathrm{crit}}\approx 0.13\hbar \omega_c$, above which the total SE 
state remains separable for all times displayed in the figure [blue 
(light gray) region]. 

Interestingly, in the vicinity of the critical temperature,  we see oscillations 
as a function of time between entangled and separable regions. To better understand
the appearance of the critical temperature and the oscillations, let us note that
for our choice of spectral density we can perform the integrals in Eq.~(\ref{ent}). 
In this way we obtain
\be\label{T-crit}
\mathcal{E}(T,t) &=& 8\kappa\mathrm{Shi}\left[\frac{\hbar \omega_c}{k_B T}\right] 
- 4\kappa i\left\{\mathrm{Si}\left[\omega_c \left(t - \frac{i\hbar}{k_B T}\right)\right] \nonumber \right.\\
&& -\left. \mathrm{Si}\left[\omega_c \left(t + \frac{i\hbar}{k_B T}\right)\right]\right\}\\
& \approx&  8\kappa\mathrm{Shi}\left[\frac{\hbar \omega_c}{k_B T}\right]  
-\frac{2\kappa (\omega_c t)}{(\omega_c t)^2+(\frac{\hbar\omega_c}{k_BT})^2}
e^{\frac{\hbar\omega_c}{k_BT}}\sin{(\omega_c t)}, \nonumber \\
& & \qquad (\textrm{as}\,\, t\to\infty).
\nonumber
\ee
Here, Si$[x]$ is the sine integral and Shi$[x]$ is hyperbolic sine integral. 
The first term on the r.h.s. is time independent, and allows us to determine
the critical temperature. The other terms are time dependent,
and are responsible for the oscillations observed in Fig.~\ref{wc-2}.
These oscillations tend to zero as $1/t$, as can be seen from the long
time behavior displayed in Eq.~(\ref{T-crit}).

Looking at the entanglement criterion for Werner states~(\ref{ent-w}), the
critical temperature is determined from
\be
8\kappa\, \mathrm{Shi}\left[\frac{\hbar \omega_c}{k_B T_{\mathrm{crit}}}\right] = \ln\left[\frac{1+c}{2c}\right],
\ee
which gives a numerical value of $k_B T_{\mathrm{crit}}\approx 0.1345\hbar\omega_c$ 
(see Fig.~\ref{wc-2}). This expression shows the dependence of the
critical temperature on the initial two-qubit state.

\begin{figure}[h]
\includegraphics[scale=0.34]{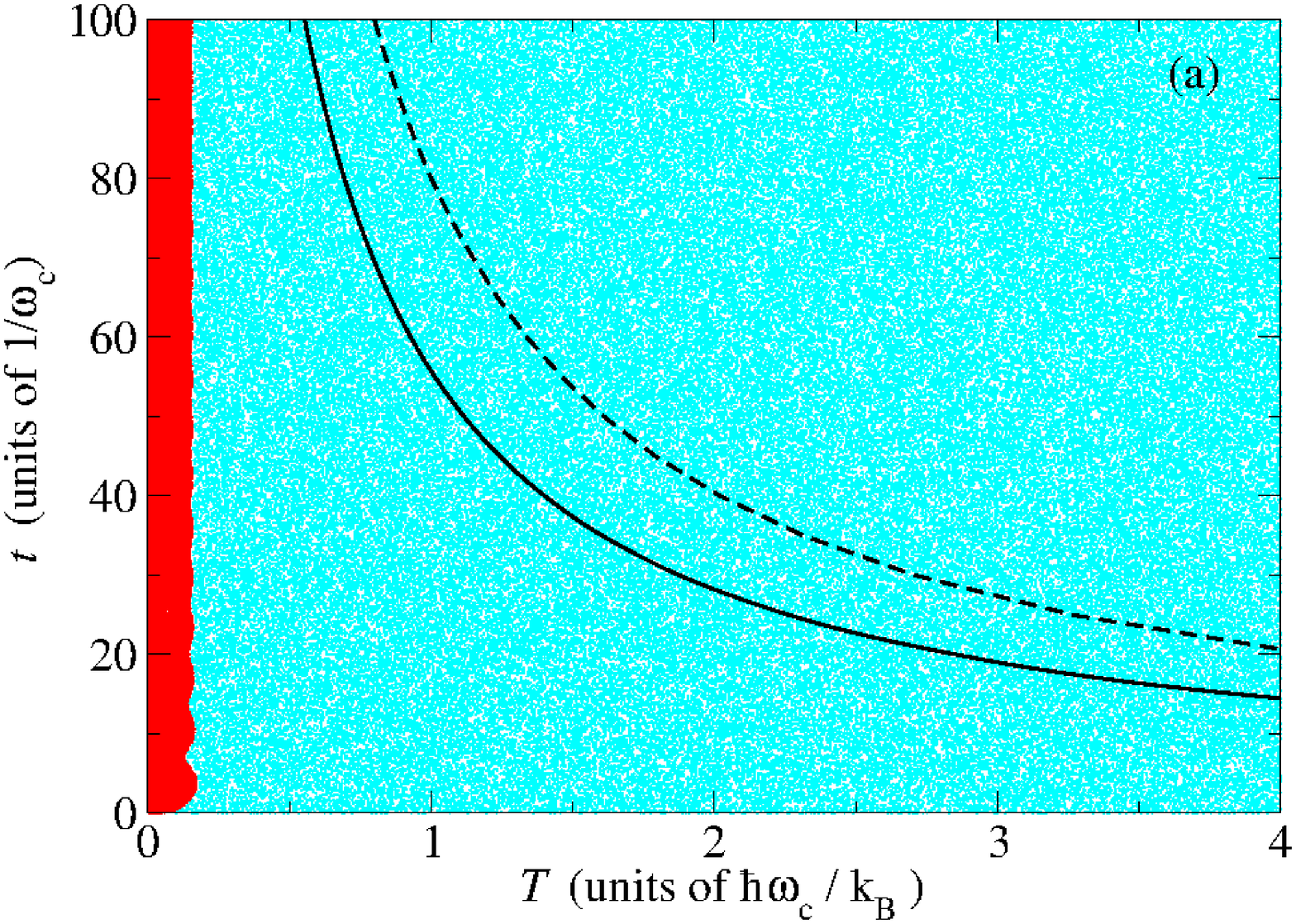}\\ 
\includegraphics[scale=0.34]{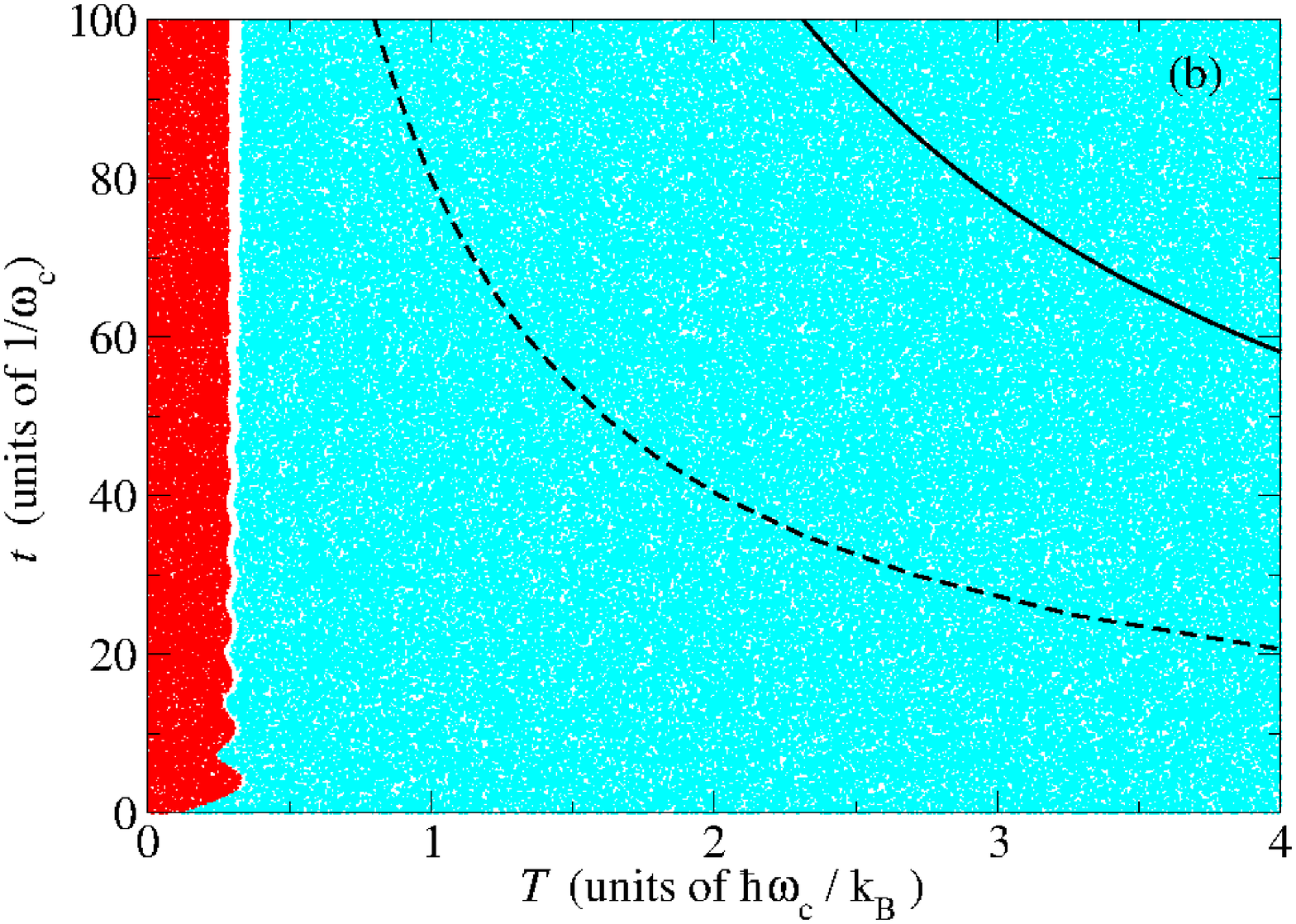}
\caption{(Color online) Temperature-time diagram. Red (dark gray) region:
 entangled SE state, blue (light gray) region: separable SE state. The 
 black dashed line indicates the decoherence time scale and the black full line 
 the two-qubit sudden death. While below the black full line the 
 two-qubit system  has some amount of entanglement, above this line the
 entanglement vanishes. Parameters are $\kappa = 10^{-3}$ (weak coupling) and (a)
 $c=0.5$ and (b) $c=0.9$ for $k_BT_{\mathrm{crit}}\approx 0.16\hbar \omega_c$
 in (a) and  for $k_BT_{\mathrm{crit}}\approx 0.29\hbar \omega_c$ in (b), the total
 state oscillates as a function of time between separable and entangled regions.}
 \label{wc-5-9} 
 \end{figure}

In Fig.~\ref{wc-5-9} we analyze the case when the two qubits are initially 
entangled [$c=0.5$ in  Fig.~\ref{wc-5-9} (a) and $c=0.9$ in Fig.~\ref{wc-5-9} (b)]. 
Again, we see SE entanglement building up for very low temperature, and 
observe a separable total state for larger temperatures and all times displayed in  
the figures. The black full line indicates, for a given temperature, the time of 
complete loss of two-qubit entanglement ({\it sudden death}). Accordingly, the 
black dashed line indicates the decoherence time scale of the two-qubit state. 
The critical temperatures for SE-entanglement are (a)\,$k_BT_{\mathrm{crit}}
\approx 0.16\hbar \omega_c$ and (b)\,$k_BT_{\mathrm{crit}}\approx 0.29\hbar 
\omega_c$, which means that the total SE state remains separable (besides the
small oscillations) for all displayed times and $T>T_{\mathrm{crit}}$.

Interestingly, our results show that the loss of entanglement between the two 
qubits does not have a direct relation with the build-up of SE-entanglement.
Unless the temperature is extremely low, we see that the decay of entanglement 
(and also decoherence) in the system happens while the SE bipartition is still in a 
separable state. While for the highly entangled, rather pure state ($c=0.9$) 
the decoherence time scale is shorter than the time for sudden death, this 
situation reverses for a less entangled initial state ($c=0.5$).

\subsection{\sl Strong coupling ($\kappa = 1$)}

Here we choose the same three Werner initial states as before. The most
significant difference to the weak 
coupling case is that our criteria for separability and entanglement no longer 
cover the whole temperature-time diagram. We observe the appearance of a white 
region, where the separability and entanglement conditions are not sufficient to 
decide whether the system is entangled with the environment or whether the total 
state is still separable.

\begin{figure}[htb] 
\includegraphics[scale=0.34]{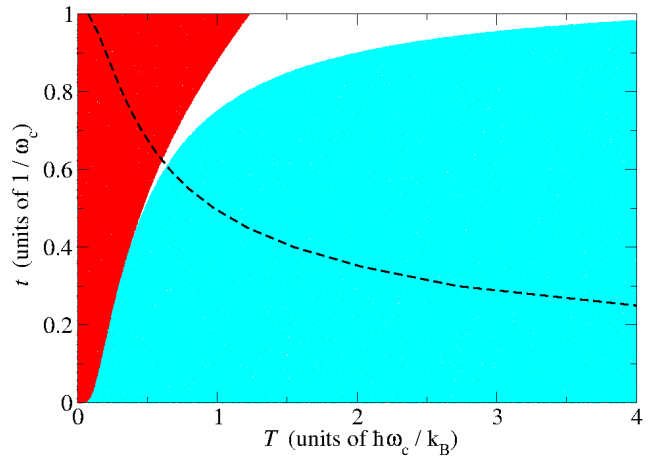} 
\caption{(Color online) Temperature-time diagram. Red (dark gray) region: 
entangled SE state, blue (light gray) region: separable SE state. Parameters 
are $\kappa = 1$ (strong coupling) and $c=0.2$ (no initial entanglement between 
the two qubits). The black dashed line indicates the decoherence time scale} 
\label{sc-2}
\end{figure}

For the initial Werner state with parameter $c=0.2$ (no entanglement between 
the qubits), we can see in Fig.~\ref{sc-2} that our criteria for entanglement and 
separability have the same border line for short times ($\omega_c t \lesssim 0.5$). 
For times $\omega_c t \gtrsim 0.5$ a gap between the separability and entanglement
conditions appear, as mentioned above. Even without any entanglement between the
qubits, due to SE interaction and the coherences of the coupled qubit, entanglement
between two-qubit system and environment builds up.

\begin{figure}[htb] 
\includegraphics[scale=0.34]{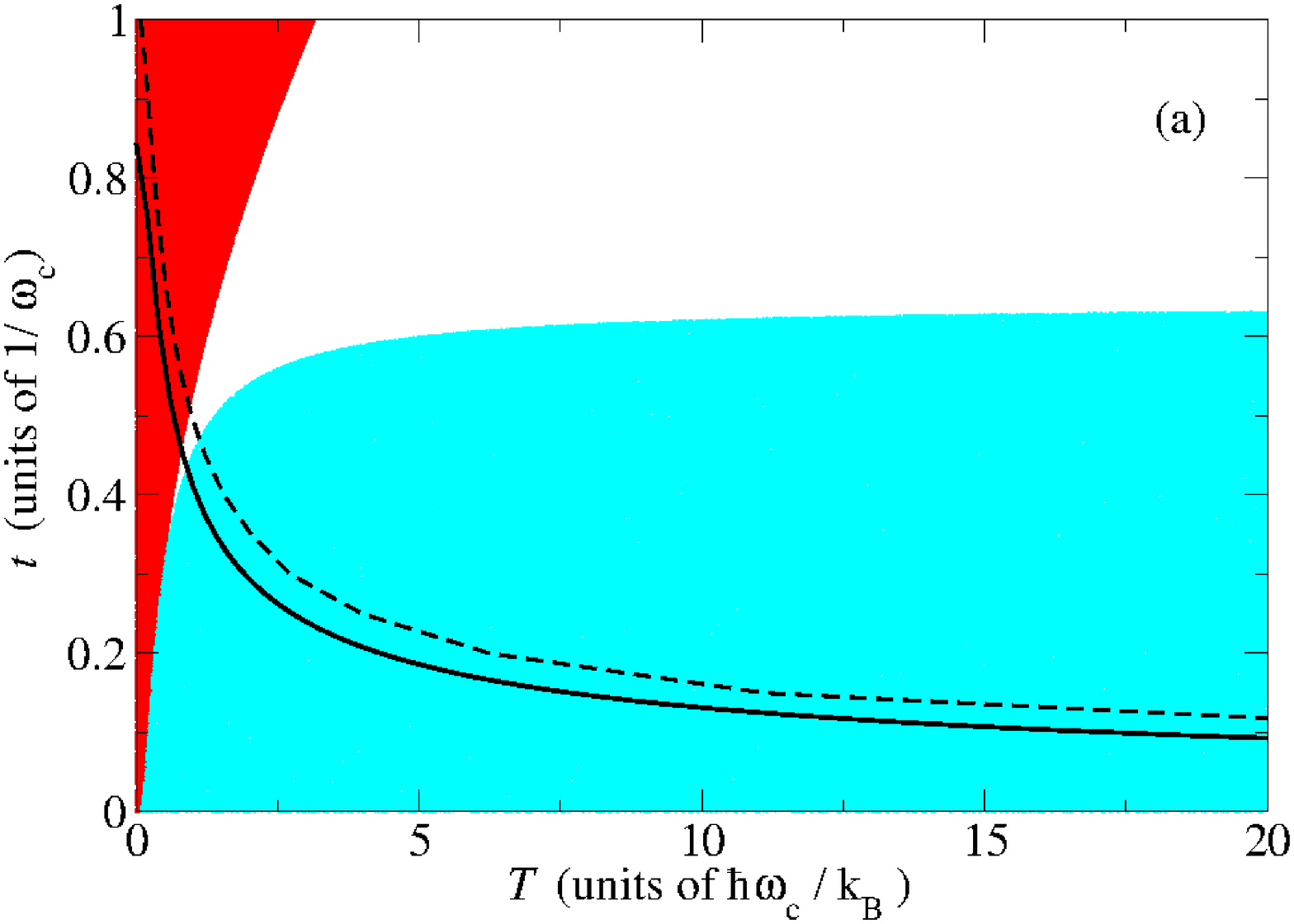}\\ 
\includegraphics[scale=0.34]{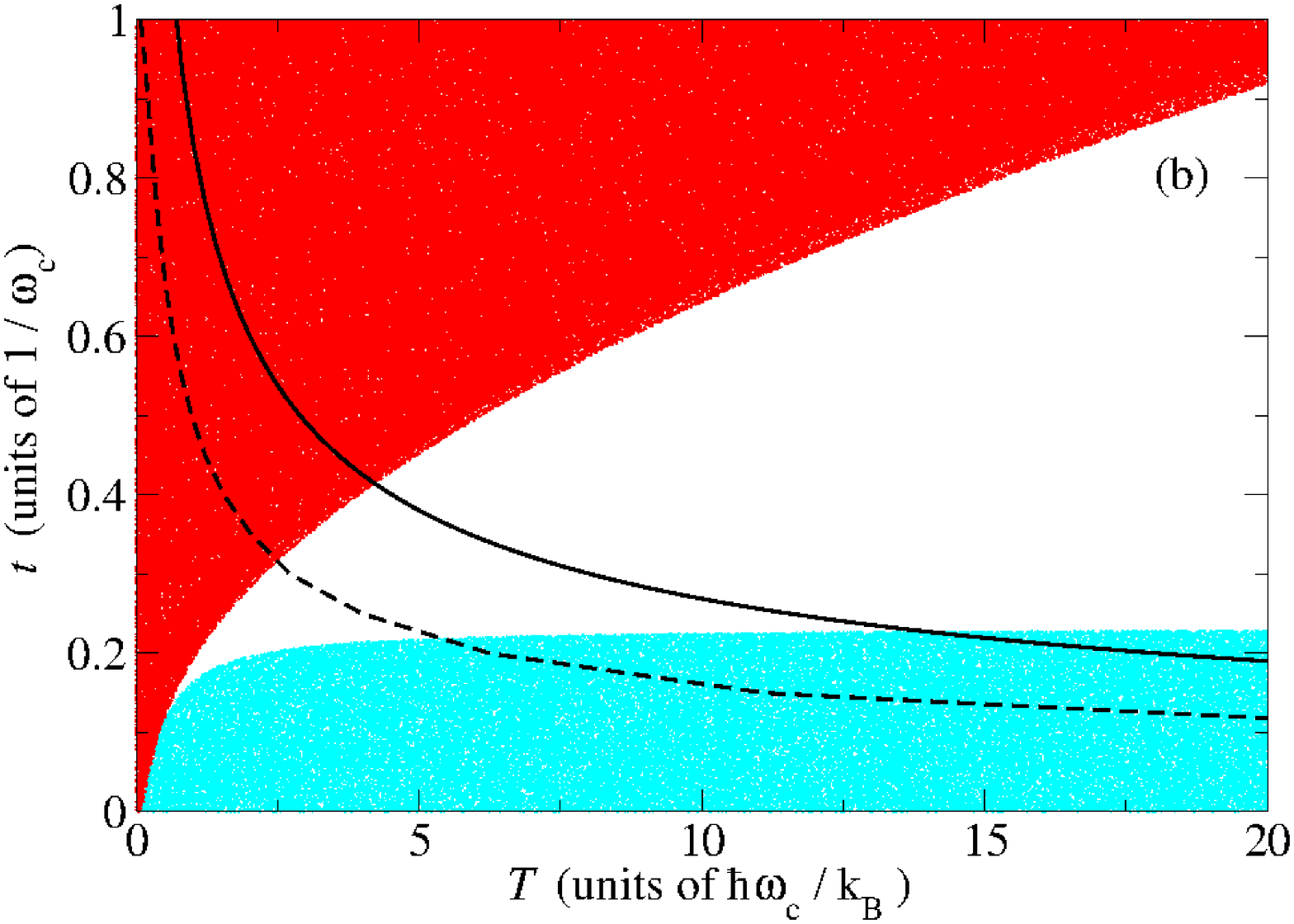} 
\caption{(Color online) Temperature-time diagram. Red (dark gray) region:
 entangled SE state, blue (light gray) region: separable SE state. The 
 black dashed line indicates the decoherence time scale and the black full line 
 the two-qubit sudden death. While below the black full line the 
two-qubit system has some amount of entanglement, above this line the entanglement
 vanishes. Parameters are $\kappa = 1$ (strong coupling) and (a) $c=0.5$ and 
 (b) $c=0.9$.} 
\label{sc-5-9}
\end{figure}

In a more interesting scenario, in Fig.~\ref{sc-5-9} we show the initial 
Werner states with parameters (a)\,$c=0.5$ and (b)\,$c=0.9$. In these cases we have 
initial entanglement between the qubits, until entanglement sudden death (black 
full line). As in the case of weak coupling, looking at the figures it is clear that 
the build-up of SE entanglement does not have any relation with the sudden
death of entanglement within the two-qubit states. 

For low temperatures, for example, there is 
still entanglement between the qubits, while the SE is first separable and then 
entangled. For larger temperatures we can see entanglement sudden death, when 
SE is still separable, and becomes entangled after some time.

In the context of strong coupling we are also able to calculate 
the  critical temperatures from Eq.~(\ref{T-crit}).
These temperatures are $k_BT_{\mathrm{crit}}\approx 7.29\hbar \omega_c$ for
$c=0.2$, $k_BT_{\mathrm{crit}}\approx 19.7\hbar \omega_c$ for $c=0.5$ and
$k_BT_{\mathrm{crit}}\approx 148\hbar \omega_c$ for $c=0.9$, and lie 
outside the temperature ranges displayed in the figures.

\subsection{Decay of concurrence}

We point out that oscillations in time between entangled and separable 
total state, as observed in Fig.~\ref{wc-2} for $T\approx T_{\mathrm{crit}}$,
are reflected to some extent in an oscillatory decay of concurrence 
in the two-qubit state. However, care has to be taken since the same type of 
oscillations in the concurrence decay are visible in regions of SE entanglement, 
or as SE separability, respectively.

\begin{figure}[htb]
\includegraphics[scale=0.33]{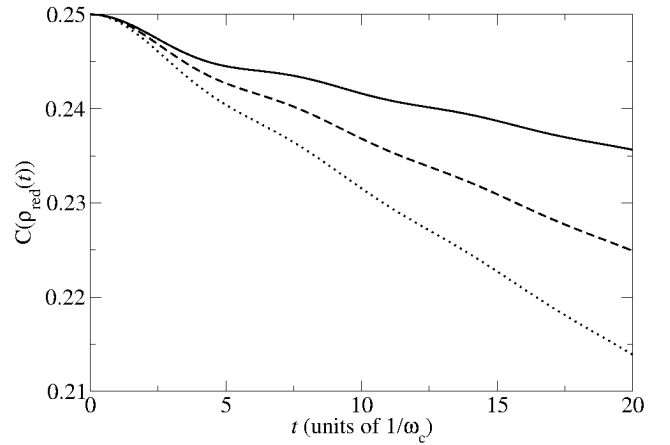}\\
\caption{Concurrence for weak coupling ($\kappa = 10^{-3}$), with $c=0.5$ and $k_B 
T = 0.1\hbar \omega_c$ (black full line), $k_B T = 0.2\hbar \omega_c$ (black dashed 
line) and $k_B T = 0.3\hbar \omega_c$ (black dotted line). We can see 
oscillations of concurrence which gets smoothed out with increase of temperature.}
\label{fig-conc} 
\end{figure}

In Fig.~\ref{fig-conc} an example of this oscillatory behavior is shown. 
For an initial Werner state with $c=0.5$ and weak coupling, we see this oscillatory 
decay of concurrence~(\ref{x_con}) due to the coupling of qubit $A$ with the 
environment. While the dashed curve corresponds to oscillations between entangled 
and separable total state, the full and dotted curves correspond to entangled and 
separable SE states, respectively. This behavior 
is more accentuated for small temperatures, and becomes smooth for larger 
temperatures.

\section{Other bipartitions}
\label{bipartitions}
So far we focused on SE entanglement and its relation to the loss of 
entanglement within the initial system ($\rho_{AB}$) state. Other authors have studied
entanglement in reduced bipartitions \cite{Maziero10, Farias12, Aguilar14} for a 
pure environmental initial state and found no bipartite entanglement in any of the 
states $\rho_{AE}=$tr$_B\{\rho_{ABE}\}$ or $\rho_{BE}=$tr$_A\{\rho_{ABE}\}$.
These findings also hold in our case, where a thermal environmental initial state
is used.

Tracing over the spectator qubit $B$, $\rho_{AE}=$tr$_B\{\rho_{ABE}\}$,
there is never any entanglement between $A$ and the $A$-environment. This is 
because we have chosen X-states whose reduced $A$-state is a diagonal mixture in 
the dephasing basis and initial separability is preserved.

Tracing over qubit $A$, the state $\rho_{BE}=$tr$_A\{\rho_{ABE}\}$ will never 
develop any entanglement between $B$ and the $A$-environment. Again this is due to 
the fact that there is no initial entanglement and $B$ is the spectator-qubit.

In \cite{Farias12,Aguilar14} the authors investigate tangle for the three-qubit ABE-state, 
showing GHZ-type three-qubit entanglement. Let us stress, however, that for a 
mixed environmental initial state the three-qubit picture ceases to hold and 
statements about multipartite entanglement are difficult to obtain.
Nevertheless, we see in Fig.~\ref{sc-5-9} that for long enough times there is entanglement
between the two qubits and environment, while all three bipartite states $\rho_{AB},
\rho_{AE}, \rho_{BE},$ are separable, pointing at genuine three partite entanglement.
 
In~\cite{Maziero10}, the authors couple a system of two qubits (X-state initially 
entangled) to two independent environments ($\rho_{ABE_AE_B}$), which they assume
to be qubits. By looking at the dynamics of correlations for different bipartitions,
they show that there does not exist any relation between the build-up of SE correlation
and the decay of entanglement in the main system. However, they do not study
the partition $AB-E$, where we find the build-up of SE (quantum) correlations.

In the present work and others discussed above~\cite{Maziero10, Farias12, Aguilar14},
the main message is that apparently there is no simple relation between the loss of 
entanglement within the system and the build-up of entanglement with the environment.
We cannot identify a {\it transfer} of entanglement from the
system to the SE partition. 

Can the lost entanglement of the system be found within the environment? 
The answer is no, as can be easily checked in our model.

Tracing over the qubits we find the $P$-representation of the environmental
state

\be
\rho_\mathrm{E} &=& \textrm{Tr}_{\mathrm{sys}}[\rho_{\mathrm{tot}}(t)] 
= \int \frac{d^2 z}{\pi} \frac{1}{\bar n}e^{-|z|^2/\bar n} \hat P_\mathrm{E}(t;z,z^*) \ket{z}\bra{z},
\nonumber
\ee
with the positive
\be
\hat P_\mathrm{E}(t;z,z^*) = 
\big[\mathcal{A}_+(\rho_{11}+\rho_{22}) + 
 \mathcal{A}_-(\rho_{33}+\rho_{44})\big]>0.
\nonumber
\ee

From the positivity of $\hat P_E(t;z,z^*)$ we can conclude that any multimode reduced 
state of the environment is a classical mixture of coherent states and thus separable.
In particular, any bipartite state of any two modes $\lambda_1, \lambda_2$ picked 
from the environment, is separable. To conclude, there is never any build-up of 
entanglement between the modes of the environment, indicating that indeed the
entanglement between the qubits may completely disappear.

\section{Conclusions}
\label{conclusions}

The understanding of system-environment correlations is fundamental 
for preserving quantum information. In this work we investigate the dynamics 
of quantum correlations (entanglement) and study how they are redistributed (or not) among 
distinct partitions.  

Any investigation of system-environment correlation starts with the choice of
the full system-environment model. Therefore, we first discuss the issue of
dilating qubit dephasing, highlighting physically different models
which lead to the same dephasing dynamics on the reduced level.

Here, we use a realistic, finite temperature, infinite degrees of freedom environment.
We investigate the entanglement dynamics of a two-qubit system coupled to a bath 
of harmonic oscillators. Using a partial $P$-representation we analyzed the exact total state
of an appropriate dephasing model. 
We derive conditions that enable us, for a large range of parameters, to detect whether 
the two-qubit--environment state is separable or entangled. Different 
coupling strengths and two-qubit initial states are considered.

The total system displays an interesting behavior when we look at the entanglement
relation between the parts. Entanglement between the two-qubit system $AB$ and
the environment $E$ will appear for low enough temperatures. For temperatures
above a critical temperature, the total state remains separable. 
For most temperatures, and during relevant times for decoherence and loss of 
entanglement in the two-qubit state, the total state remains separable. 
Additionally, we show that the modes of the environment never get entangled.  
Therefore, for most parameters, the initial entanglement between the two qubits
vanishes without any build-up of entanglement in any other bipartition.
At very low temperatures we detect conditions where the total state oscillates
between separable and entangled domains as a function of time. Remarkably, these
oscillations can also be seen in the time evolution of the entanglement between the
two qubits.

The investigation of SE separability and entanglement for the highly
nontrivial total state, rests on a partial $P$ representation. For 
entanglement detection we have to guess a good test state. This
approach leads to ``white" regions in the $(T,t)$ diagram, 
where we cannot make any statement about whether the total state
is entangled or separable. Moreover, our approach allows one to detect SE 
entanglement, but  we do not quantify it. Consequently, do not look at 
monogamy relations in the  tripartite $ABE$ system. A search for a way to 
estimate  the amount of entanglement and verify these relations is still an 
open issue.

\section*{Acknowledgments}

MWB thanks CNPq and ACSC thanks CAPES and the National Institute for Science and Technology 
of Quantum Information (INCT-IQ) for financial support. WTS would like to thank Luiz
Davidovich and his group for a stimulating discussion on entanglement dynamics in open systems.
The Curitiba-Dresden collaboration is made possible through support from the 
PROBRAL joint Brazilian-German program of CAPES and DAAD.


\end{document}